\documentclass[pdflatex,sn-mathphys-ay]{sn-jnl}

\usepackage{graphicx}
\usepackage{subcaption}
\usepackage{amsmath,amssymb,amsfonts}
\usepackage{bm}
\usepackage{mathtools}
\usepackage{booktabs}
\usepackage{array}
\usepackage{siunitx}
\usepackage{xcolor}
\usepackage{float}
\usepackage{multirow}
\usepackage{comment}
\usepackage{lineno}

\newcommand{\dd}{\,\mathrm{d}}
\newcommand{\E}{\mathbb{E}}
\newcommand{\Lop}{\mathcal{L}}
\newcommand{\Pop}{\mathcal{P}}

\newcommand{\Mop}{\mathcal{M}}
\newcommand{\JJA}{\mathrm{JJA}}
\newcommand{\Med}{\mathrm{Med}}
\newcommand{\TXx}{\mathrm{TXx}}
\newcommand{\TXp}{\mathrm{TX90p}}
\newcommand{\WSDI}{\mathrm{WSDI}}
\newcommand{\RMSE}{\mathrm{RMSE}}

\begin{document}

\title[Projected memory and European warm-tail risk]{Projected climate memory and inherited warm-tail risk in accelerated European summer warming}

\author*[1]{\fnm{Mauricio} \sur{Herrera-Mar\'in}}
\email{mherrera@udd.cl}
\author[1]{\fnm{Alex}      \sur{Godoy-Fa\'undez}}
\author[1]{\fnm{Diego}     \sur{Rivera}}

\affil*[1]{%
	\orgdiv{Research Center on Sustainability and Strategic Resource
		Management (CISGER), Faculty of Engineering},
	\orgname{Universidad del Desarrollo},
	\orgaddress{\street{Avda.~Plaza 700}, \city{Santiago}},
	\country{Chile}}

\abstract{
	European summer warming reflects the interaction of smooth background change,
	persistent ocean--land--circulation states, and same-season atmospheric and
	land-surface innovation. We develop an empirical reduced-dynamics framework in
	which regional summer indicators are decomposed into inherited slow-state
	memory, the component of the slow state predictable from previous years, and a
	contemporaneous innovation. Projection-operator theory provides the conceptual
	motivation for this decomposition: reduced climate observables are expected to
	carry memory of omitted or only partially observed degrees of freedom. The
	statistical implementation is deliberately operational and data-constrained,
	using finite causal memory filters, ridge-regularised state prediction, and
	logistic event-risk models rather than attempting to identify a unique continuous
	memory kernel from short annual records.
	
	Using ERA5-derived annual summer indicators for 28 IPCC AR6 European
	sub-regions over 1950--2024, with temporal validation over 2006--2024, we find
	that Mediterranean-state memory improves mean summer-temperature prediction
	relative to trend-only and ARX baselines. The incremental gain over ARX is
	modest, and moving-average, exponentially weighted, and tempered filters often
	contain similar annual information, indicating that the data identify useful
	slow-state memory more robustly than a unique kernel shape. The predictable-state
	decomposition is diagnostically informative but sensitive to ridge
	regularisation; the reconstructed current state is therefore treated as an
	identity-based diagnostic rather than as a forecasting model.
	
	The strongest result concerns warm-tail risk. Under a parsimonious linear
	logistic risk model, high accumulated Mediterranean memory increases predicted
	upper-tail event probability by about 8--11 percentage points for annual maximum
	summer temperature, warm-day frequency, and warm-spell duration at 1-, 3-, and
	5-year horizons. The same model underestimates the larger observed high-minus-low
	memory differences, suggesting that accumulated memory loads the system into a
	more susceptible regime while same-season circulation and land-surface
	conditions help determine event realisation. Regional bootstrap intervals remain
	positive for all targets and horizons. Circular-shift placebos, which preserve
	memory autocorrelation while breaking chronological alignment, yield one-sided
	probabilities of approximately 0.05--0.14 and do not survive strict family-wise
	correction across the nine related tests. The evidence is therefore moderate
	rather than decisive. Overall, annual projected memory is not a universal
	short-horizon predictor; it is a physically interpretable inherited
	risk-loading variable that helps identify years and regions predisposed to
	warm-tail outcomes.
}


\maketitle

\section{Introduction}\label{sec:introduction}

Europe is a hotspot of recent warming and heat-extreme intensification. Observational and reanalysis studies show rapid increases in European summer temperature, heatwaves and warm-tail indicators, with strong regional contrasts and important roles for circulation persistence, jet-stream configurations, blocking, soil-moisture feedbacks, Mediterranean warming and land--atmosphere coupling \citep{Rousi2022,Vautard2023,Patterson2023,Seneviratne2010,Hirschi2011,Miralles2014,Pfahl2014,Schaller2018,Holmberg2023}. Persistent double-jet states have been linked to accelerated western European heatwave trends \citep{Rousi2022}, and the hottest days in north-western Europe can warm faster than mean summer days \citep{Patterson2023,Vautard2023}. These findings suggest that European warming is not adequately described as a smooth trend plus independent weather noise. It is shaped by slow reservoirs, persistent atmospheric regimes and contemporaneous innovations.

A common element in this literature is persistence. Heatwaves persist; blocking and jet-stream states can persist; soil moisture carries information from previous precipitation and evapotranspiration anomalies; and regional sea-surface temperatures provide slower thermal reservoirs. Yet persistence is often represented empirically through autocorrelation, residence times, lagged predictors or event durations. These quantities are useful, but they do not by themselves separate three distinct objects: (i) information inherited from previous slow states, (ii) the part of the current state predictable from that history, and (iii) the same-year innovation that realises or suppresses an extreme. This distinction is important because a model can predict the conditional mean well while still failing to indicate whether a warm-tail event occurred on top of an inherited high-risk state.

This paper focuses on that distinction. We use the term \emph{projected climate memory} in a deliberately modest sense: a regional observable is treated as a reduced description of a higher-dimensional climate system, and lagged slow-state histories are used as empirical summaries of the information inherited from omitted or only partially observed degrees of freedom. Projection-operator theory, particularly the Mori--Zwanzig identity \citep{Mori1965,Zwanzig1973,ChorinHald2013}, provides a useful conceptual reason to expect memory in reduced descriptions. However, the empirical analysis below does not estimate a Mori--Zwanzig kernel, solve a generalized Langevin equation or identify a full Volterra series. The operational models are finite causal filters, ridge-regularised autoregressive distributed-lag regressions and logistic event-risk models. This explicit separation between conceptual motivation and statistical implementation is essential for the present study.

The contribution is empirical and diagnostic. First, we ask whether lagged slow-state histories, especially Mediterranean thermal memory, improve annual summer-temperature prediction relative to trend-only and autoregressive baselines. Second, we decompose slow states into predictable and innovation components. Third, we test whether accumulated memory amplifies the probability of warm-tail and warm-spell events. The central hypothesis is not that memory must dominate every short-horizon mean forecast. It is that accumulated slow-state memory acts as an inherited risk-loading variable: it shifts the background susceptibility of the system, while same-season atmospheric and land-surface innovations help determine whether an extreme is realised.

We apply this framework to ERA5-derived annual summer indicators for 28 IPCC AR6 European sub-regions over 1950--2024 \citep{Hersbach2020,Iturbide2020}. The targets are mean summer temperature $T_{\JJA}$, annual maximum summer temperature $\TXx_{\JJA}$, warm-day frequency $\TXp_{\JJA}$ and warm-spell duration $\WSDI_{\JJA}$. Candidate slow states include Mediterranean thermal state, regional dryness, Z500 and blocking occurrence. All main predictive comparisons use temporal validation over 2006--2024. The paper is written for the applied climate-dynamics community: the theoretical language motivates why memory is plausible, but the claims are restricted to what the empirical tests support.

\section{Conceptual framework}\label{sec:framework}

\begin{figure}[htbp]
\centering
\includegraphics[width=0.96\textwidth]{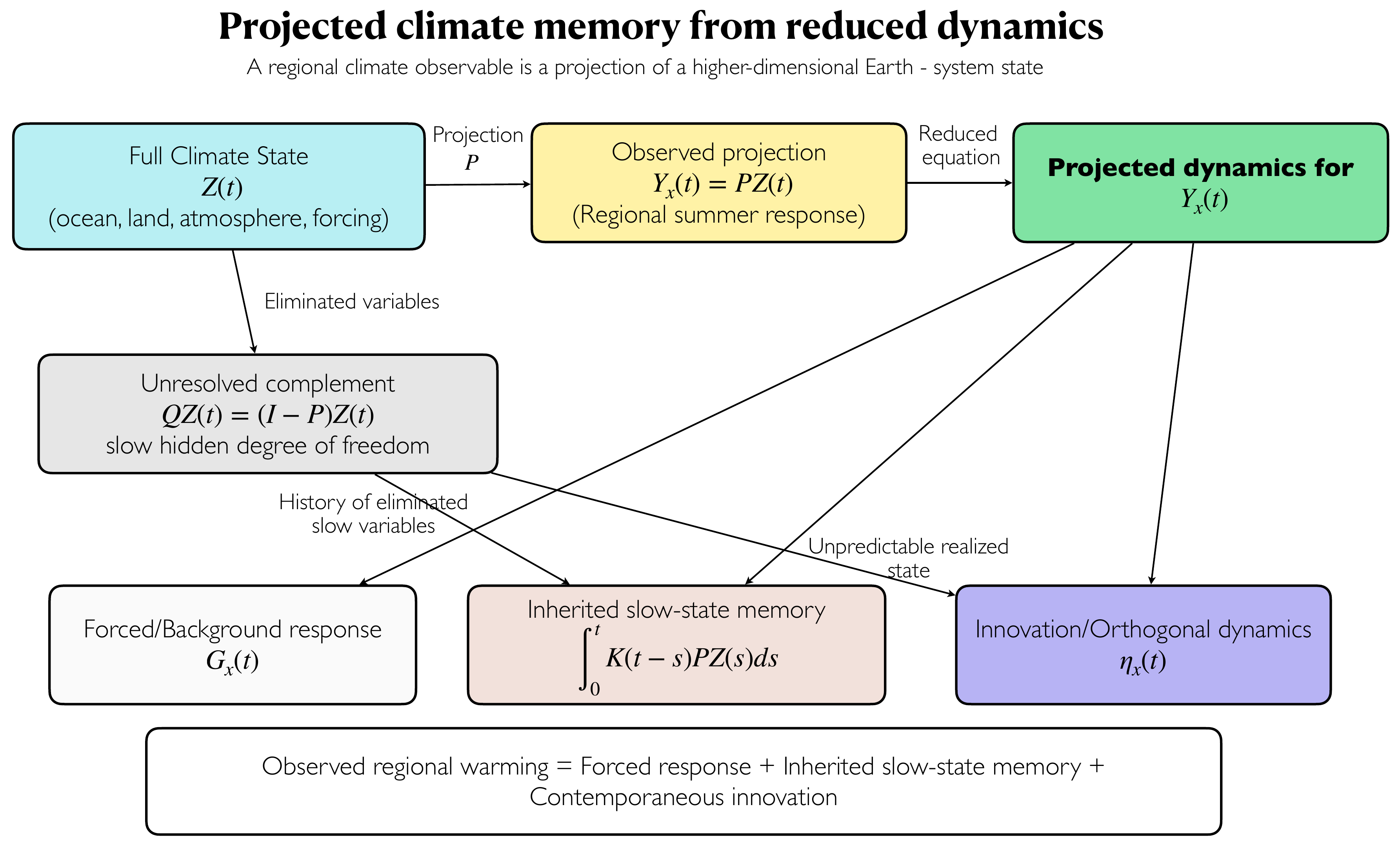}
\caption{Conceptual framework. A regional summer climate indicator is treated as a reduced observable of a coupled ocean--land--atmosphere system. The empirical analysis separates lagged slow-state memory, the component of the slow state predictable from previous years and the same-year innovation. The diagram is conceptual and should not be read as an estimated Mori--Zwanzig model.}
\label{fig:framework}
\end{figure}

Let $Z(t)$ denote the full climate state and let $Y_x(t)$ be a resolved regional observable, such as summer temperature in region $x$. Projection formalisms show that an exact reduced equation for $Y_x(t)$ generally contains an instantaneous tendency, a memory integral and an orthogonal or innovation term. In schematic form,
\begin{equation}
\frac{\dd}{\dd t}\Pop Z(t)=\Pop\Lop\Pop Z(t)+\int_0^t K(t-s)\Pop Z(s)\,\dd s+\eta(t),
\label{eq:mz_schematic}
\end{equation}
where $\Pop$ is a projection, $K$ is a memory kernel and $\eta(t)$ is an unresolved-dynamics term. Equation~\eqref{eq:mz_schematic} is used here as a motivation, not as an estimated model. The empirical counterpart is the decomposition of an observed slow state $S_x(t)$ into a part predictable from past information and a contemporaneous innovation:
\begin{equation}
S_x(t)=\widehat{S}_x(t\mid t-1)+\varepsilon_x(t).
\label{eq:state_decomp}
\end{equation}
This equation is not a claim that $\varepsilon_x(t)$ is the exact Mori--Zwanzig orthogonal dynamics. It is an operational diagnostic: $\widehat{S}_x(t\mid t-1)$ is the component that can be inferred from lagged annual information, while $\varepsilon_x(t)$ is the residual current-year component. Figure~\ref{fig:framework} summarises this interpretation.

For a slow driver $Q_x(t)$, the causal memory filters used below have the finite form
\begin{equation}
\Mop_Q(t)=\sum_{\tau=1}^{K} w(\tau) Q_x(t-\tau), \qquad \sum_{\tau=1}^{K}w(\tau)=1.
\label{eq:memory_filter}
\end{equation}
We compare moving-average, exponentially weighted and tempered weights. The tempered family is
\begin{equation}
w_{\alpha,\theta,K}(\tau)=\frac{(\tau+1/2)^{-\alpha}\exp(-\theta\tau)}{\sum_{j=1}^{K}(j+1/2)^{-\alpha}\exp(-\theta j)},
\label{eq:tempered_weights}
\end{equation}
with $0<\alpha<1$ and $\theta>0$. These filters are first-order causal lag filters. They are not claimed to identify a unique physical kernel. Similar performance of moving-average, EWMA and tempered filters is interpreted as evidence that annual data contain finite slow-state memory but do not uniquely identify its functional shape.

\section{Data and methods}\label{sec:data_methods}

\subsection{Data and variables}\label{subsec:data}

The analysis uses ERA5-derived annual summer indicators for 28 IPCC AR6 European sub-regions over 1950--2024 \citep{Hersbach2020,Iturbide2020}. The response variables are $T_{\JJA}$, $\TXx_{\JJA}$, $\TXp_{\JJA}$ and $\WSDI_{\JJA}$. Candidate slow states are Mediterranean thermal state $A_{\Med}$, regional dryness $D_x$, Z500 and blocking occurrence. Models are calibrated over 1950--2005 and evaluated over 2006--2024. All predictors are standardised using the calibration period.

The term $G(t)$ denotes a smooth trend-control variable. In the present analysis the available state table does not contain a complete physically based radiative-forcing series, so $G(t)$ is implemented as standardised calendar year. It is therefore a trend-only control for smooth background warming, not a causal estimate of radiative forcing from greenhouse gases, aerosols or solar variability. For this reason, all comparisons below refer to a ``trend-only'' baseline. The results should not be interpreted as formal attribution relative to a complete external-forcing counterfactual.

\subsection{Mean-response models}\label{subsec:mean_models}

The mean-response hierarchy compares a trend-only baseline, an autoregressive baseline and causal memory filters:
\begin{align}
B_0:\quad Y_x(t)&=\mu_x+\beta_xG(t-1)+e_x(t),\\
B_{\mathrm{ARX}}:\quad Y_x(t)&=\mu_x+\beta_xG(t-1)+\sum_{\tau=1}^{p}\phi_{\tau,x}Y_x(t-\tau)+e_x(t),\\
M:\quad Y_x(t)&=\mu_x+\beta_xG(t-1)+\gamma_x\Mop_Q(t)+e_x(t).
\end{align}
The memory model is implemented with moving-average, EWMA and tempered filters. The goal is to test whether lagged slow-state histories contain information beyond trend and autoregressive persistence, not to prove that one filter is universally optimal.

\subsection{Predictable state and innovation}\label{subsec:state_model}

The predictable component of a slow state is estimated with a region-specific ARDL--ridge model,
\begin{align}
\widehat{S}_x(t\mid t-1)
&=\alpha_x+
\sum_{\ell=1}^{3}a_{\ell,x}S_x(t-\ell)+
\sum_{\ell=1}^{3}b_{\ell,x}T_{\JJA,x}(t-\ell)+
\sum_{\ell=1}^{3}c_{\ell,x}G(t-\ell) \nonumber\\
&\quad+
\sum_{Q\in\mathcal{S}\setminus\{S\}}\sum_{\ell=1}^{3}d_{Q,\ell,x}Q_x(t-\ell),
\label{eq:state_ardl}
\end{align}
where $\mathcal{S}=\{A_{\Med},D,Z500,B\}$ is the set of available slow states. Equation~\eqref{eq:state_ardl} is deliberately a statistical state-prediction model. Its residual is the innovation in Eq.~\eqref{eq:state_decomp}. The penalty parameter is evaluated over $\lambda\in\{0.01,0.1,1,10,100\}$ because the annual sample is short relative to the lagged feature set.

When $\widehat{S}_x(t\mid t-1)$ and $\varepsilon_x(t)$ are added together, they reconstruct the current state $S_x(t)$. This reconstruction is a diagnostic upper bound on the information contained in the realised state; it is not a causal forecast. We therefore do not present the reconstructed-state RMSE as a predictive achievement.

\subsection{Tail-risk amplification and placebo tests}\label{subsec:tail_methods}

For each target and region, an upper-tail event is defined by the calibration-period 90th percentile,
\begin{equation}
E_x^{(h)}(t)=\mathbf{1}\{Y_x(t+h)>q_{0.90,x}^{\mathrm{train}}\}, \qquad h\in\{1,3,5\}.
\end{equation}
The main event-risk model is logistic regression with trend control and accumulated Mediterranean memory,
\begin{equation}
\Pr\{E_x^{(h)}(t)=1\}=\sigma\left(\beta_{0,x}^{(h)}+\beta_{1,x}^{(h)}G(t-1)+\beta_{2,x}^{(h)}\Mop_{A_{\Med}}(t)\right).
\label{eq:event_model}
\end{equation}
The tail-amplification index is
\begin{equation}
\Delta p_{\mathrm{tail}}=
\E\{p(E=1\mid M\in Q_{0.8})\}-
\E\{p(E=1\mid M\in Q_{0.2})\},
\label{eq:tail_amp}
\end{equation}
where $M=\Mop_{A_{\Med}}$, and $Q_{0.2}$ and $Q_{0.8}$ are the lower and upper memory quintiles within the validation period. We report both predicted amplification, based on fitted probabilities, and observed amplification, based on empirical event frequencies.

Uncertainty is assessed by regional bootstrap. Chronology dependence is tested using circular-shift placebos: within each region the memory series is shifted by a random non-zero offset, preserving its marginal distribution and autocorrelation while breaking alignment with subsequent events. Nine related placebo tests are performed (three targets and three horizons). We therefore report the unadjusted one-sided placebo probabilities and the false-discovery-rate implication; the tests are treated as supportive diagnostics rather than definitive significance claims.

\section{Results}\label{sec:results}

\subsection{Lagged slow-state memory helps mean summer temperature, but kernel shape is not unique}\label{subsec:mean_results}

\begin{figure}[htbp]
\centering
\includegraphics[width=0.98\textwidth]{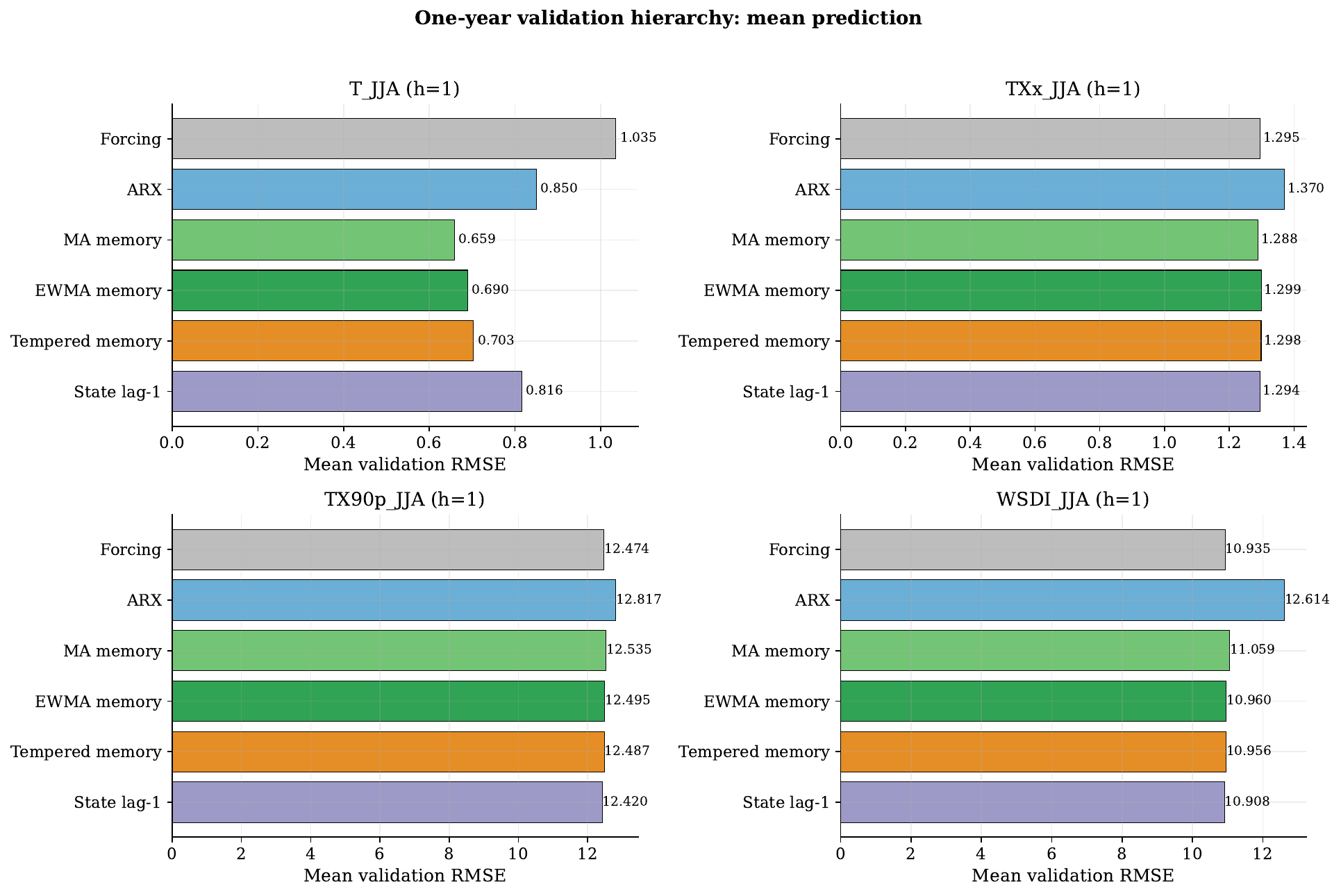}
\caption{One-year validation hierarchy. Memory filters are compared with trend-only and autoregressive baselines. The figure should be interpreted as a hierarchy of empirical information sources rather than a proof of a unique memory kernel.}
\label{fig:model_hierarchy}
\end{figure}

For $T_{\JJA}$, Mediterranean-state memory improves validation skill relative to trend-only and ARX baselines (Table~\ref{tab:tjja_hierarchy}; Fig.~\ref{fig:model_hierarchy}). However, MA, EWMA and the tempered filter perform similarly. This is an important result rather than a weakness: annual data identify useful slow-state memory more robustly than they identify a unique kernel shape.

\begin{table}[htbp]
\centering
\caption{Representative validation hierarchy for mean summer temperature. Values are mean validation RMSE across regions for 2006--2024. The reconstructed current-state row is a diagnostic upper bound, not a causal forecast.}
\label{tab:tjja_hierarchy}
\begin{tabular}{lcl}
\toprule
Model class & Mean RMSE & Interpretation \\
\midrule
Trend-only baseline & 0.847 & Smooth annual trend control \\
Persistence / random-walk baseline & 0.922 & Local one-year persistence \\
ARX baseline & 0.767 & Autoregressive plus trend information \\
Distributed lag & 0.700 & Flexible finite memory \\
Tempered causal memory filter & 0.696 & Structured slow-state memory \\
Predictable slow-state component & 0.693 & Strictly lagged state prediction \\
Reconstructed current state & 0.448 & Diagnostic contemporaneous upper bound \\
\bottomrule
\end{tabular}
\end{table}

\begin{figure}[htbp]
\centering
\includegraphics[width=0.96\textwidth]{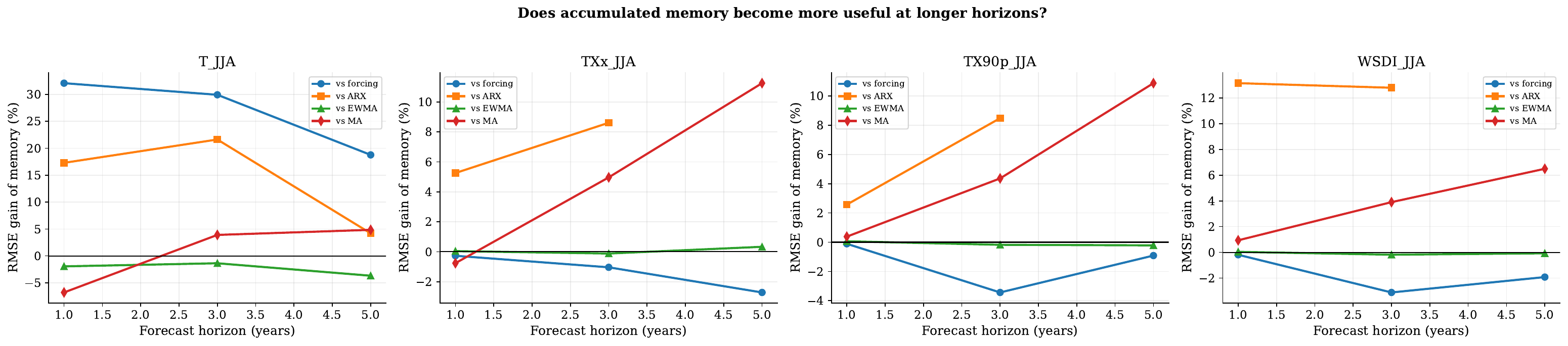}
\caption{Horizon-dependent memory contribution. Values show RMSE gain of the tempered Mediterranean memory filter relative to selected baselines at 1-, 3- and 5-year horizons. Positive values indicate lower RMSE for the memory model.}
\label{fig:horizon_gain}
\end{figure}

At longer horizons, memory gains are most evident for mean summer temperature and weaker or target-dependent for warm-tail indicators (Fig.~\ref{fig:horizon_gain}; Table~\ref{tab:horizon_gains}). The tempered Mediterranean-memory filter reduces RMSE for $T_{\JJA}$ by about 32\%, 30\% and 19\% relative to the trend-only baseline at 1-, 3- and 5-year horizons, respectively, and by about 17\%, 22\% and 4\% relative to ARX. For warm-tail indicators, the same memory filter is not uniformly better than the trend-only or state-lag baselines, although it often improves over ARX where ARX estimates are available. This supports a focused interpretation: annual slow-state memory contributes most clearly to mean summer temperature, while its strongest role for extremes is not mean-RMSE reduction but tail-risk loading.

\begin{table}[htbp]
\centering
\caption{Multi-horizon RMSE of the tempered Mediterranean-memory filter and relative gains. Gains are computed as $100\,(\RMSE_{\rm baseline}-\RMSE_{\rm memory})/\RMSE_{\rm baseline}$. Positive values favour memory. Dashes indicate that the corresponding ARX estimate was not retained in the validation summary.}
\label{tab:horizon_gains}
\begin{tabular}{llrrrr}
\toprule
Target & Horizon & Memory RMSE & Gain vs trend & Gain vs ARX & Best model RMSE \\
\midrule
$T_{\JJA}$ & 1 yr & 0.703 & 32.1\% & 17.3\% & 0.659 \\
$T_{\JJA}$ & 3 yr & 0.679 & 29.9\% & 21.6\% & 0.670 \\
$T_{\JJA}$ & 5 yr & 0.753 & 18.8\% & 4.2\% & 0.726 \\
$\TXx_{\JJA}$ & 1 yr & 1.298 & -0.3\% & 5.3\% & 1.288 \\
$\TXx_{\JJA}$ & 3 yr & 1.308 & -1.0\% & 8.6\% & 1.295 \\
$\TXx_{\JJA}$ & 5 yr & 1.330 & -2.7\% & -- & 1.295 \\
$\TXp_{\JJA}$ & 1 yr & 12.487 & -0.1\% & 2.6\% & 12.420 \\
$\TXp_{\JJA}$ & 3 yr & 12.903 & -3.4\% & 8.5\% & 12.474 \\
$\TXp_{\JJA}$ & 5 yr & 12.588 & -0.9\% & -- & 12.431 \\
$\WSDI_{\JJA}$ & 1 yr & 10.956 & -0.2\% & 13.2\% & 10.908 \\
$\WSDI_{\JJA}$ & 3 yr & 11.276 & -3.1\% & 12.8\% & 10.935 \\
$\WSDI_{\JJA}$ & 5 yr & 11.145 & -1.9\% & -- & 10.935 \\
\bottomrule
\end{tabular}
\end{table}

\subsection{The state-prediction model needs regularisation}\label{subsec:ridge_results}

\begin{figure}[htbp]
\centering
\includegraphics[width=0.76\textwidth]{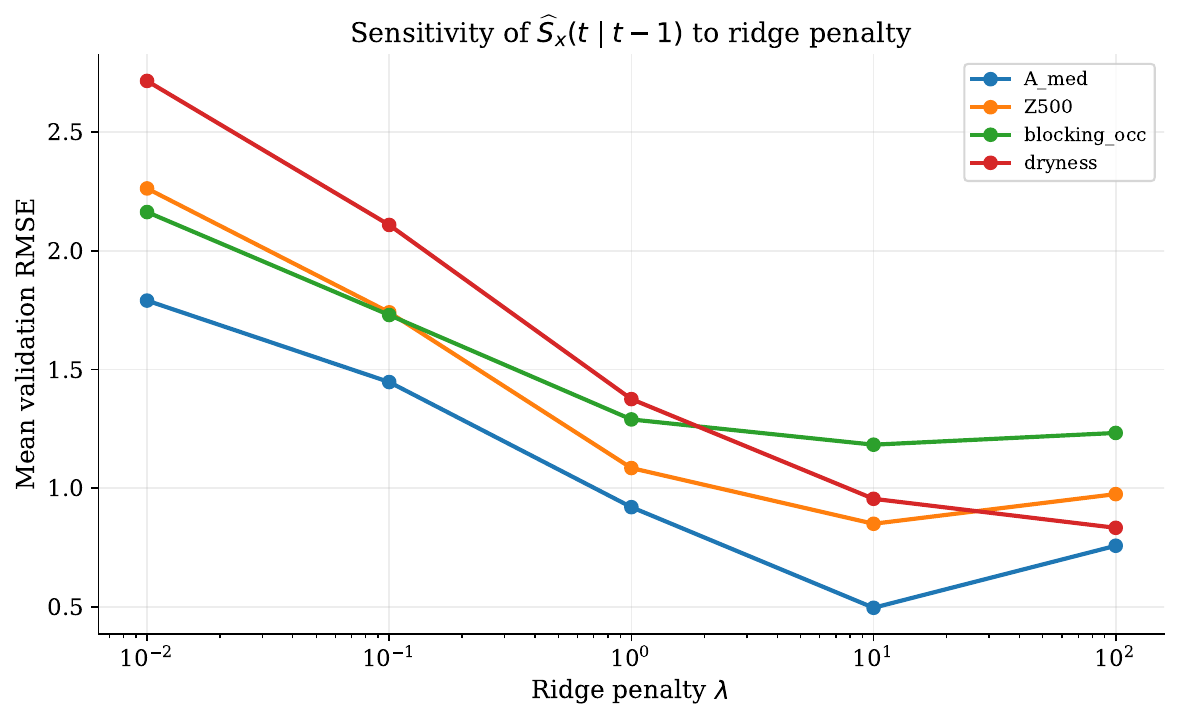}
\caption{Sensitivity of the predictable-state model to ridge regularisation. Small penalties under-regularise the ARDL model, while stronger penalties stabilise out-of-sample state prediction.}
\label{fig:ridge}
\end{figure}

The ARDL--ridge model for $\widehat{S}_x(t\mid t-1)$ is sensitive to under-regularisation. Stronger penalties improve out-of-sample prediction of slow states, especially for $A_{\Med}$ and dryness (Table~\ref{tab:ridge}; Fig.~\ref{fig:ridge}). This confirms that the predictable-state model should be treated as a stabilised diagnostic model, not as an unconstrained high-dimensional regression.

\begin{table}[htbp]
\centering
\caption{Ridge-penalty sensitivity for the predictable-state model. Values report mean validation RMSE across regions for representative penalties.}
\label{tab:ridge}
\begin{tabular}{lccc}
\toprule
Slow state & RMSE at $\lambda=1$ & Best reported penalty & RMSE at best reported penalty \\
\midrule
$A_{\Med}$ & 0.920 & 10 & 0.497 \\
Z500 & 1.085 & 10 & 0.850 \\
Blocking occurrence & 1.290 & 10 & 1.184 \\
Dryness & 1.376 & 100 & 0.833 \\
\bottomrule
\end{tabular}
\end{table}

\begin{figure}[htbp]
\centering
\includegraphics[width=0.90\textwidth]{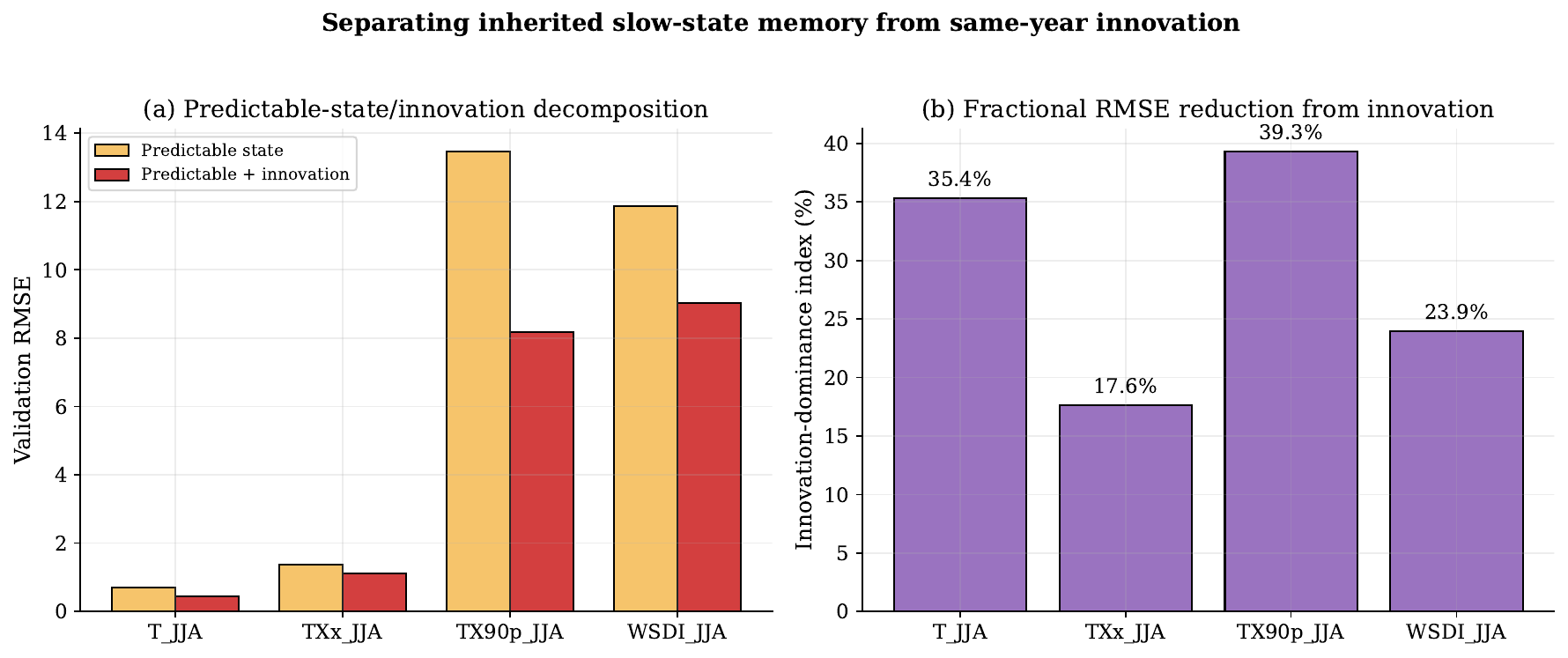}
\caption{Predictable-state and innovation decomposition. The predictable component is estimated from lagged information; the innovation is the same-year residual component. The reconstructed state is a diagnostic representation of realised conditions, not a causal forecast.}
\label{fig:state_decomp}
\end{figure}

The predictable-state decomposition itself is shown in Fig.~\ref{fig:state_decomp}. The low error obtained by reconstructing the current state from predictable and innovation components is expected from the identity in Eq.~\eqref{eq:state_decomp}; it is not used as evidence that memory alone forecasts the realised current state.

\subsection{Memory and innovation are coupled, but correlations are diagnostic}\label{subsec:coupling_results}

\begin{figure}[htbp]
\centering
\includegraphics[width=0.68\textwidth]{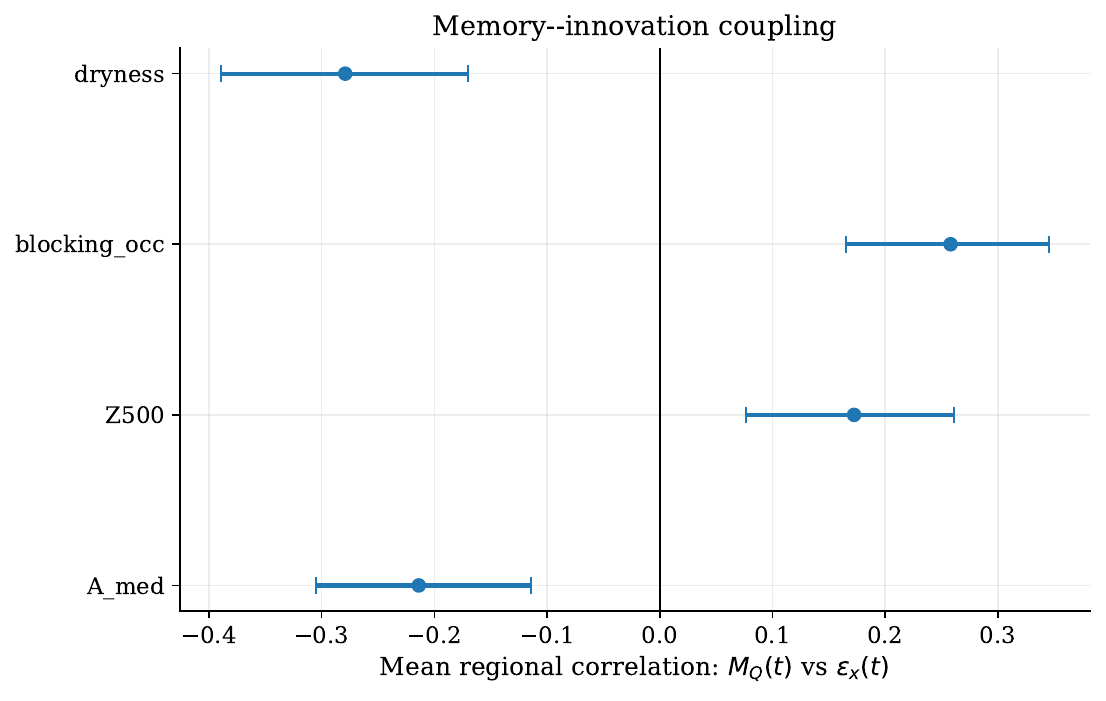}
\caption{Memory--innovation association. Points show regional mean correlations between accumulated memory and ARDL innovation; bars indicate regional-bootstrap intervals. The result indicates coupling between inherited loading and current-year residual state, not exact orthogonality.}
\label{fig:meminnov}
\end{figure}

Accumulated memory and same-year innovation are not independent (Table~\ref{tab:meminnov}; Fig.~\ref{fig:meminnov}). Mediterranean and dryness memory have negative mean association with their innovations, consistent with partial mean reversion after high accumulated states. Z500 and blocking memory have positive association, consistent with regime persistence. These correlations are moderate and should be interpreted diagnostically; they are residual associations from the ARDL model, not direct estimates of the Mori--Zwanzig orthogonal term.

\begin{table}[htbp]
\centering
\caption{Association between accumulated memory and same-year innovation. Values are regional mean correlations with 95\% regional-bootstrap intervals.}
\label{tab:meminnov}
\begin{tabular}{lcc}
\toprule
State & Mean correlation & 95\% CI \\
\midrule
$A_{\Med}$ & -0.214 & [-0.305, -0.114] \\
Dryness & -0.279 & [-0.389, -0.170] \\
Z500 & 0.172 & [0.076, 0.261] \\
Blocking occurrence & 0.258 & [0.165, 0.345] \\
\bottomrule
\end{tabular}
\end{table}

\subsection{Accumulated Mediterranean memory amplifies warm-tail risk}\label{subsec:tail_results}

\begin{figure}[htbp]
\centering
\includegraphics[width=0.98\textwidth]{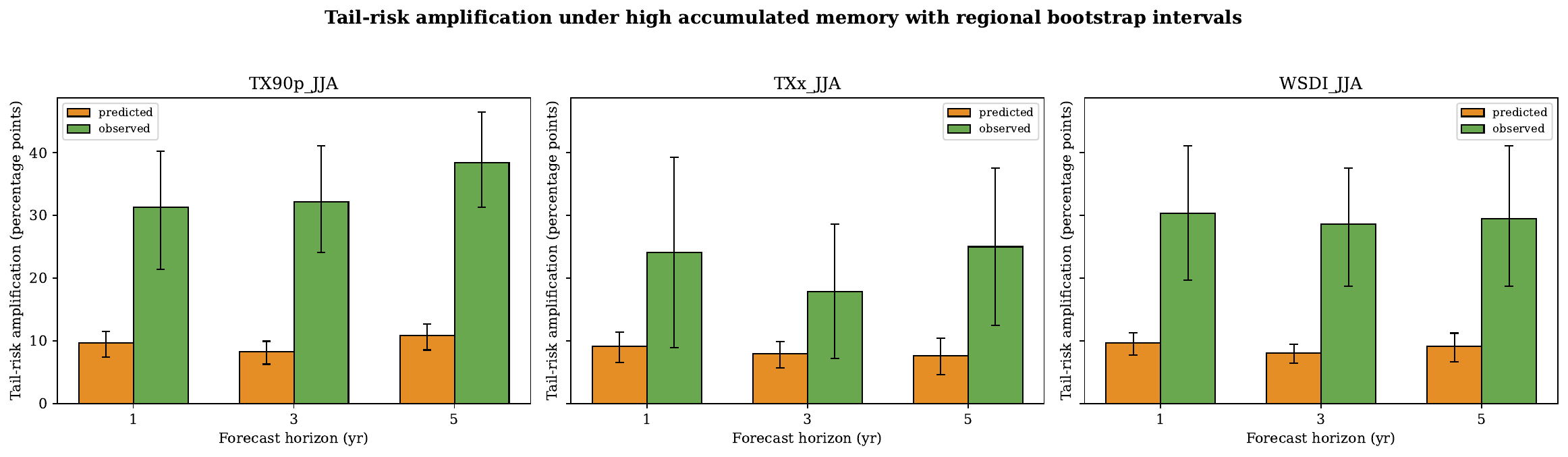}
\caption{Tail-risk amplification under high accumulated Mediterranean memory. Bars show mean predicted amplification and regional-bootstrap intervals for warm-tail targets and forecast horizons.}
\label{fig:tail_boot}
\end{figure}

The strongest empirical result is the tail-risk amplification. Across $\TXx_{\JJA}$, $\TXp_{\JJA}$ and $\WSDI_{\JJA}$, high accumulated Mediterranean memory increases predicted upper-tail event probability by approximately 8--11 percentage points at 1-, 3- and 5-year horizons (Table~\ref{tab:tailamp}; Fig.~\ref{fig:tail_boot}). Regional bootstrap intervals remain positive in all cases. Observed high-minus-low memory differences are larger, approximately 18--38 percentage points, showing that the linear logistic model captures the direction of the effect but underestimates its empirical magnitude.

\begin{table}[htbp]
\centering
\caption{Tail-risk amplification by accumulated Mediterranean memory. Predicted amplification is the high-minus-low memory difference in fitted event probability; observed amplification is the corresponding empirical event-frequency difference.}
\label{tab:tailamp}
\begin{tabular}{llcccc}
\toprule
Target & Horizon & Predicted $\Delta p_{\rm tail}$ & 95\% CI & Observed $\Delta p_{\rm tail}$ & $p_{\rm circ}$ \\
\midrule
$\TXp_{\JJA}$ & 1 yr & 0.097 & [0.074, 0.115] & 0.313 & 0.066 \\
$\TXp_{\JJA}$ & 3 yr & 0.083 & [0.063, 0.099] & 0.321 & 0.123 \\
$\TXp_{\JJA}$ & 5 yr & 0.108 & [0.085, 0.127] & 0.384 & 0.050 \\
$\TXx_{\JJA}$ & 1 yr & 0.092 & [0.066, 0.114] & 0.241 & 0.072 \\
$\TXx_{\JJA}$ & 3 yr & 0.079 & [0.057, 0.098] & 0.179 & 0.119 \\
$\TXx_{\JJA}$ & 5 yr & 0.077 & [0.045, 0.103] & 0.250 & 0.138 \\
$\WSDI_{\JJA}$ & 1 yr & 0.096 & [0.078, 0.113] & 0.304 & 0.062 \\
$\WSDI_{\JJA}$ & 3 yr & 0.081 & [0.064, 0.095] & 0.286 & 0.123 \\
$\WSDI_{\JJA}$ & 5 yr & 0.091 & [0.067, 0.112] & 0.295 & 0.098 \\
\bottomrule
\end{tabular}
\end{table}

\begin{figure}[htbp]
\centering
\includegraphics[width=1.0\textwidth]{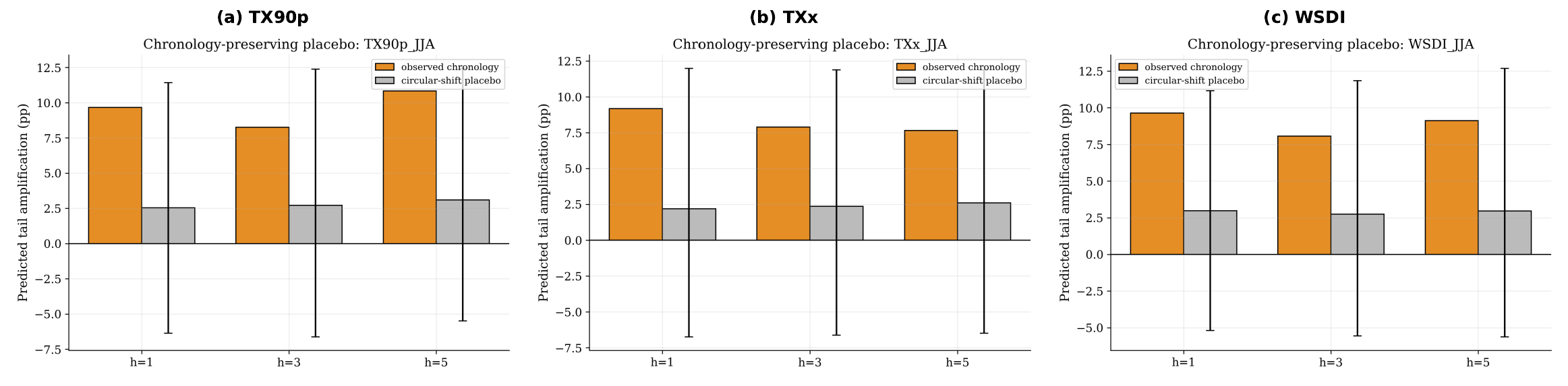}
\caption{Circular-shift chronology placebo. The placebo preserves the marginal distribution and autocorrelation of accumulated memory within each region but breaks chronological alignment with subsequent events. Observed-chronology amplification is consistently larger than the placebo mean, although the evidence is moderate under multiple-test correction.}
\label{fig:circular}
\end{figure}

Circular-shift placebo tests produce lower amplification than the observed chronology (Fig.~\ref{fig:circular}). The one-sided placebo probabilities range from 0.05 to 0.14. Applying Benjamini--Hochberg FDR correction across the nine related target--horizon tests gives a common adjusted value of approximately 0.138; a Bonferroni correction would be more conservative. We therefore interpret the placebo evidence as moderate chronological support rather than strict family-wise significance.

\subsection{Event-risk scores are useful as rankings, not calibrated probabilities}\label{subsec:calibration_results}

\begin{figure}[htbp]
	\centering
	\includegraphics[width=1.0\textwidth]{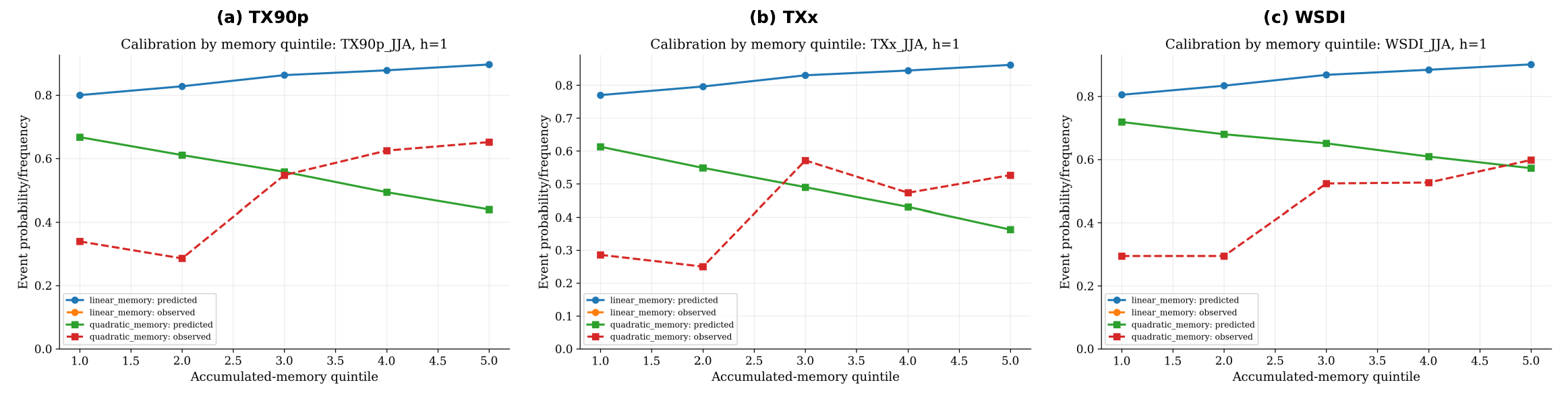}
	\caption{Calibration by accumulated-memory quintile. Predicted probabilities and observed frequencies generally increase with memory, but the logistic model overestimates the absolute probability level.}
	\label{fig:calibration}
\end{figure}

\begin{figure}[H]
	\centering
	
	\begin{subfigure}{0.95\textwidth}
		\centering
		\includegraphics[width=\linewidth]{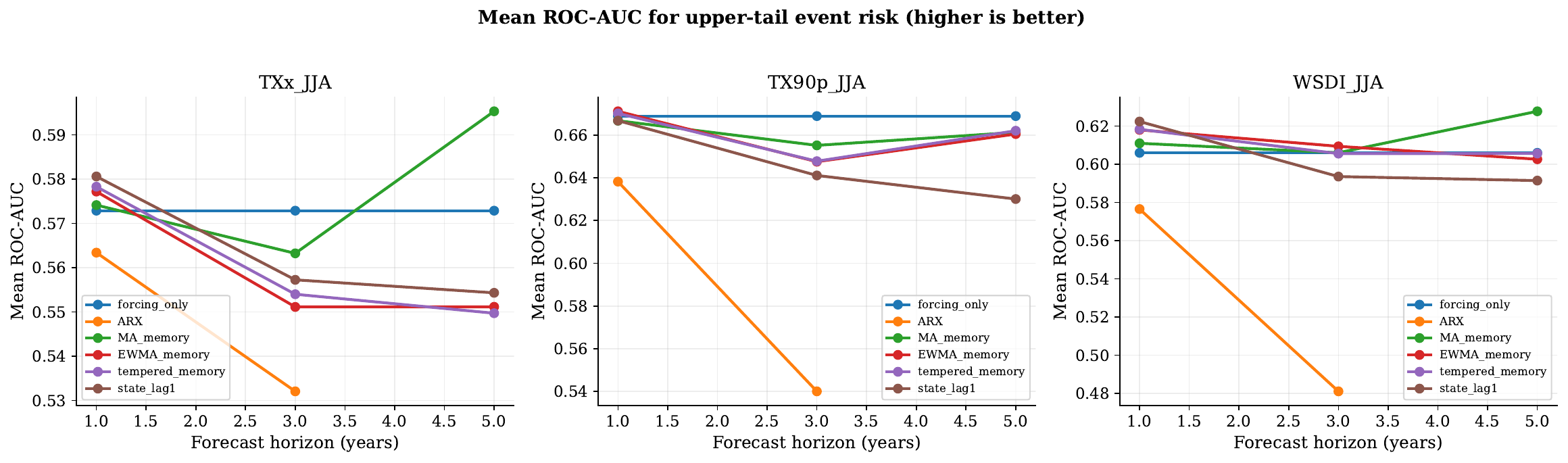}
		\caption{ROC-AUC.}
		\label{fig:risk_ranking_auc}
	\end{subfigure}
	
	\vspace{0.5cm}
	
	\begin{subfigure}{0.95\textwidth}
		\centering
		\includegraphics[width=\linewidth]{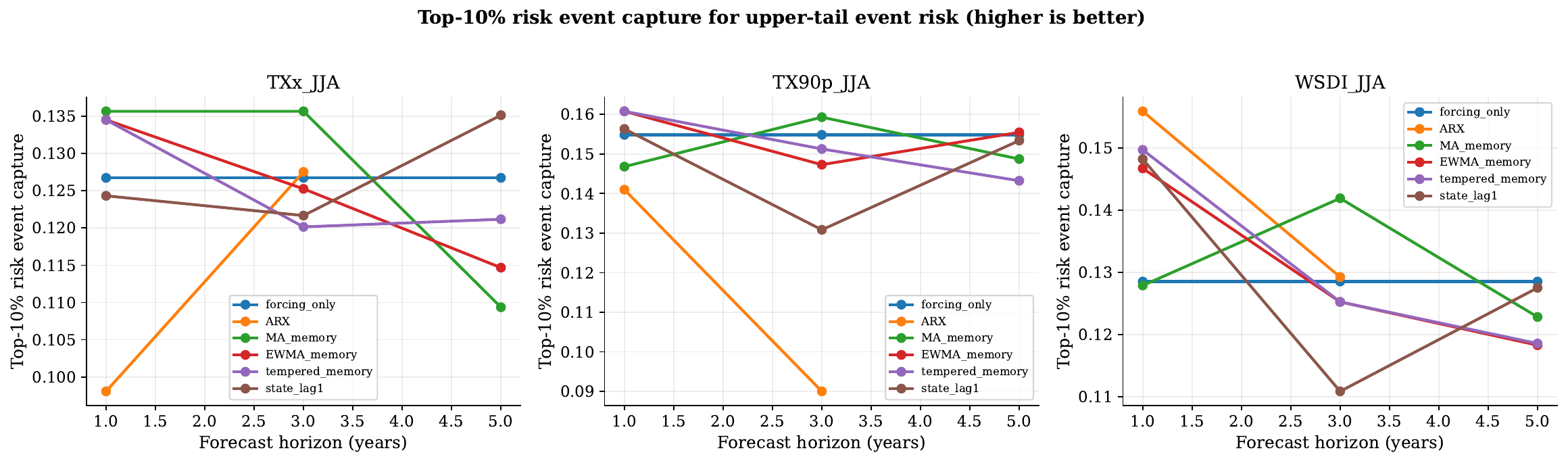}
		\caption{Top-risk capture.}
		\label{fig:risk_ranking_top10}
	\end{subfigure}
	
	\caption{Event-risk ranking diagnostics. The memory model is used primarily as a relative-risk score for identifying high-risk years, not as a fully calibrated probability model.}
	\label{fig:risk_ranking}
\end{figure}

The logistic memory model is better interpreted as a relative-risk score than as a calibrated probability model. AUC and top-risk capture indicate that memory helps rank high-risk years for several targets and horizons (Fig.~\ref{fig:risk_ranking}). Calibration by memory quintile shows that predicted and observed risk both tend to increase with memory, but the absolute probability level is overestimated (Fig.~\ref{fig:calibration}). This is why Table~\ref{tab:tailamp} emphasises differences between high- and low-memory states rather than absolute probabilities.

\begin{figure}[H] 
	\centering
	\begin{subfigure}[b]{0.95\textwidth}
		\centering
		\includegraphics[width=\textwidth]{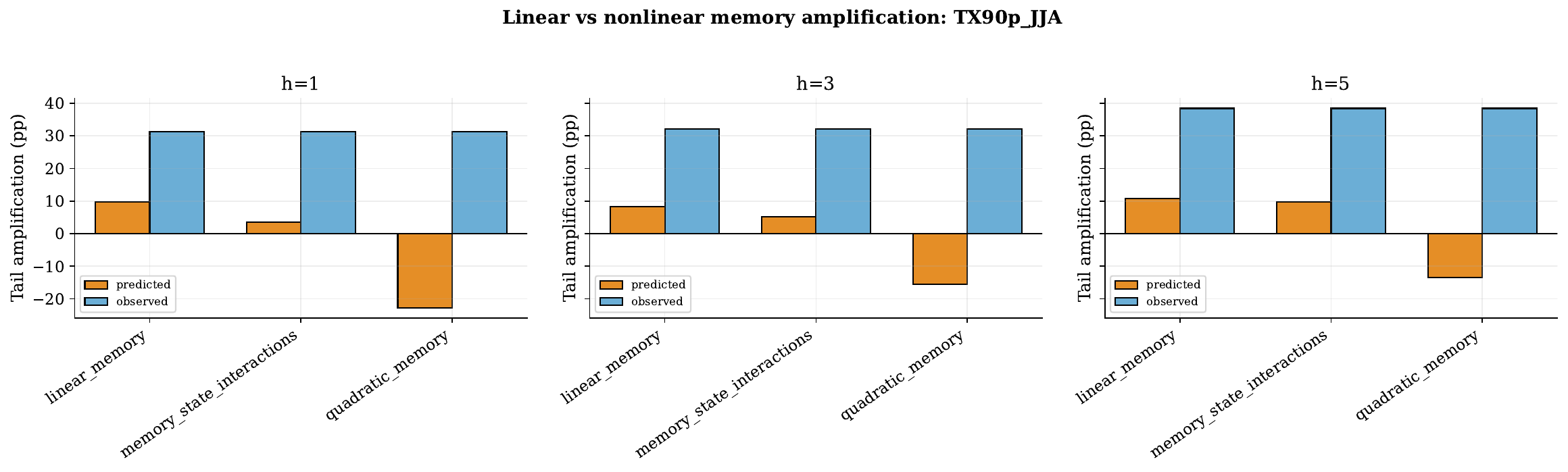}
		\caption{TX90p (Frequency of warm days)}
		\label{fig:nonlinear_tx90p}
	\end{subfigure}
	\vspace{0.3cm} 
	
	\begin{subfigure}[b]{0.95\textwidth}
		\centering
		\includegraphics[width=\textwidth]{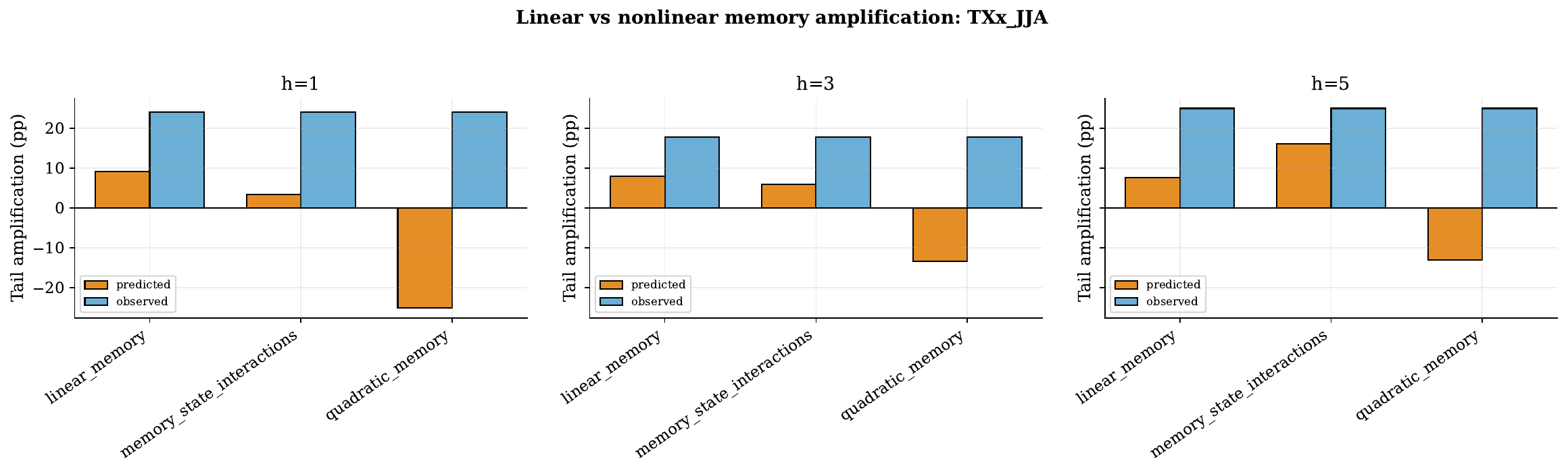}
		\caption{TXx (Maximum summer temperature)}
		\label{fig:nonlinear_txx}
	\end{subfigure}
	\vspace{0.3cm}
	
	\begin{subfigure}[b]{0.95\textwidth}
		\centering
		\includegraphics[width=\textwidth]{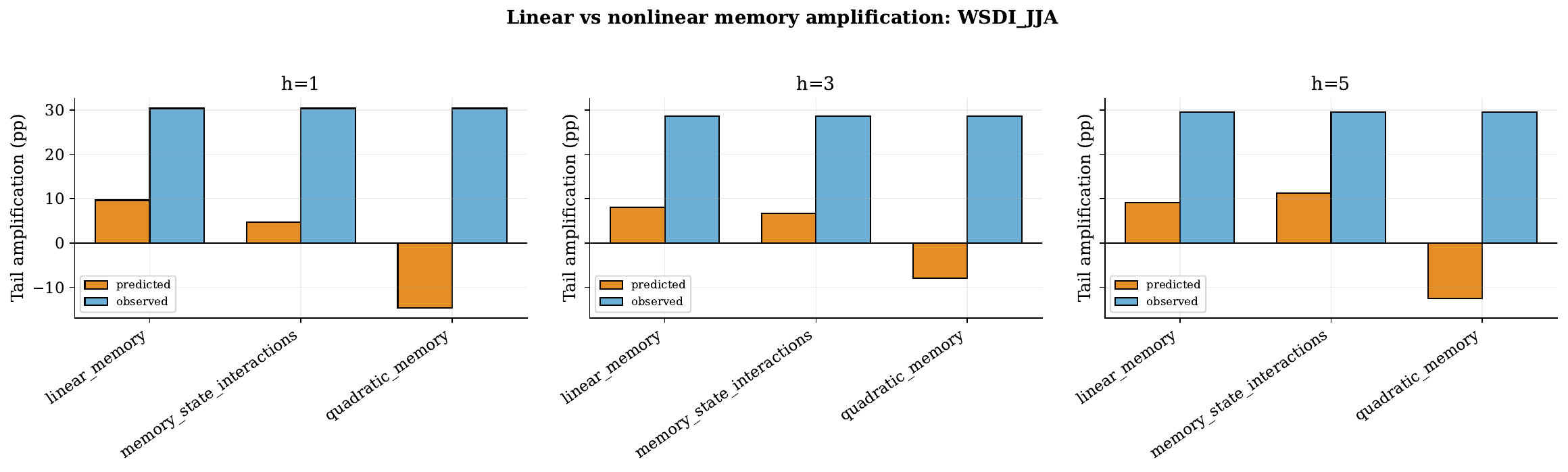}
		\caption{WSDI (Warm Spell Duration Index)}
		\label{fig:nonlinear_wsdi}
	\end{subfigure}
	
	\caption{Sensitivity analysis of tail-risk amplification across alternative non-linear and interaction event-risk specifications for summer extremes: (a) TX90p, (b) TXx, and (c) WSDI. Markers and intervals denote the estimated amplification factors and their bootstrap uncertainty under different functional forms. The unconstrained quadratic and interaction formulations exhibit severe instability and sign reversals across the short annual validation samples, whereas the linear memory model provides structural stability and conservative risk bounds.}
	\label{fig:nonlinear}
\end{figure}

The nonlinear sensitivity analysis clarifies why the main event-risk specification is kept deliberately parsimonious. Across the three warm-tail targets, the linear memory model preserves a positive predicted amplification signal at all evaluated horizons and is directionally consistent with the observed high-minus-low memory differences (Fig.~\ref{fig:nonlinear}). Adding memory--state interaction terms can change the magnitude of the predicted amplification, and in some cases increases it at longer horizons, but the effect is not uniform across targets. By contrast, the unconstrained quadratic specification is unstable in the short annual validation sample: while the observed amplification remains positive, the fitted quadratic model can reverse the sign of the predicted amplification. This sign reversal is difficult to interpret as a robust nonlinear climate mechanism and is more plausibly a finite-sample overfitting artefact. We therefore retain the linear memory model as the conservative main specification and treat interaction and quadratic forms as sensitivity checks rather than as competing operational models.

\section{Discussion}\label{sec:discussion}

\subsection{What the revised evidence supports}\label{subsec:discussion_support}

The evidence supports a focused conclusion. Annual projected memory is useful less as a universal mean predictor than as a diagnostic of inherited risk loading. Mediterranean-state memory improves mean summer-temperature prediction relative to trend and ARX baselines, but the precise kernel shape is not identifiable from annual data. For extremes, the strongest signal is not global RMSE improvement but a consistent shift in warm-tail event risk under high accumulated Mediterranean memory.

This is a narrower and more defensible claim than saying that the Mori--Zwanzig formalism has been empirically estimated. The projection viewpoint explains why memory and innovation should arise in reduced descriptions, but the reported quantities are empirical filters and residual diagnostics. The paper's contribution is therefore to connect a reduced-dynamics interpretation with transparent annual tests of inherited risk loading.

\subsection{Predictable state, innovation and contemporaneous information}\label{subsec:discussion_innovation}

The predictable-state decomposition helps avoid conflating contemporaneous physical association with causal memory. If the innovation is added back to the predictable component, the current state is reconstructed by construction. Such a model is a diagnostic upper bound on state information, not a forecast. The meaningful comparisons are those that use strictly lagged information: trend plus lagged response, trend plus causal memory, and predictable state estimated from past values. This distinction is essential when interpreting the strong role of Mediterranean or circulation states in European summer warming.

\subsection{Statistical strength and multiple testing}\label{subsec:discussion_testing}

The regional bootstrap intervals for predicted tail amplification are consistently positive, indicating robust cross-regional directionality. The circular-shift placebo is more conservative: it preserves autocorrelation and distributional structure while breaking temporal alignment. The resulting one-sided probabilities are near but not uniformly below 0.05, and do not survive strict family-wise correction across all target--horizon combinations. We therefore describe the placebo evidence as moderate. This does not erase the amplification result; rather, it calibrates its evidential strength. The most defensible interpretation is that accumulated Mediterranean memory is a reproducible risk-loading indicator, with chronology-dependent support that should be strengthened in future analyses using longer or sub-seasonal data.

\subsection{Why predicted and observed amplification differ}\label{subsec:discussion_gap}

Observed high-minus-low memory differences are larger than the probabilities predicted by the linear event-risk model. This discrepancy is informative. It suggests that high memory states may load the system into a more susceptible regime, while the realised extreme depends on same-season circulation, blocking and land-surface feedbacks. Because annual samples are short, flexible quadratic and interaction models are unstable. The gap between predicted and observed amplification should therefore be interpreted as evidence that the linear memory score captures the direction of risk loading but not the full nonlinear event-generation mechanism.

\subsection{Limitations}\label{subsec:limitations}

The study has four main limitations. First, annual data restrict the identifiability of memory kernels and nonlinear event-risk relationships. Second, the trend-control variable is standardised year in the present implementation and is not a complete radiative-forcing series; the baseline should therefore be read as smooth trend control, not causal attribution. Third, the predictable-state ARDL model is sensitive to regularisation, showing the need for parsimony in short annual samples. Fourth, circular-shift placebos provide moderate rather than decisive family-wise significance after accounting for the nine related target--horizon tests. These limitations motivate sub-seasonal extensions with explicit circulation and soil-moisture state variables and, in future work, replacement of the trend proxy by externally derived forcing estimates.

\section{Conclusions}\label{sec:conclusions}

This paper reframes annual European summer warming through the lens of inherited slow-state memory and contemporaneous innovation. The projection viewpoint motivates why reduced regional observables should contain memory, but the empirical contribution is deliberately operational: causal memory filters, predictable-state decompositions, tail-amplification indices and chronology-preserving placebos.

The main result is that accumulated Mediterranean memory acts as an inherited warm-tail risk-loading variable. It modestly improves mean summer-temperature prediction relative to trend and ARX baselines, and under the linear event-risk model it consistently increases the predicted probability of upper-tail events for $\TXx_{\JJA}$, $\TXp_{\JJA}$ and $\WSDI_{\JJA}$ at 1-, 3- and 5-year horizons. The evidence is strongest under regional bootstrap and moderate under circular-shift placebo tests after multiple-test caution. Memory should therefore not be evaluated only by short-horizon mean RMSE or by strict family-wise significance of annual placebo tests. Its value is to identify inherited susceptibility to extremes and to separate that susceptibility from same-year innovation.

\backmatter

\bmhead{Data availability}
All empirical results are derived from ERA5-based regional annual indicators and reproducible notebooks. Data and code used to construct the final figures, tables, bootstrap intervals and circular-shift placebos will be made available soon in a public repository according to journal policy.

\bmhead{Code availability}

It will be archived soon in the GitHub repository
\url{https://github.com/mauricio-herrera/europe-memory-warming}
(DOI: 10.5281/zenodo.XXXXXXX).

\bmhead{Competing interests}
The author declares no competing interests.

\bmhead{Author contributions}
M.H.-M.\ conceived and designed the study, performed all analyses
and wrote the manuscript.
A.G.-F.\ and D.R.\ contributed to interpretation and revision.

\bmhead{Acknowledgements}
ERA5, RAPID--MOCHA--WBTS and NOAA/CPC teams for publicly available data.

ANID/ANILLO ATE250004.

\bibliography{paperB_references}

\end{document}